\documentclass[
superscriptaddress,
amsmath,
amssymb,
twocolumn,
prd,
showpacs,
floatfix,
]{revtex4-2}

\usepackage{lineno}
\usepackage{color}
\usepackage{float}
\usepackage{verbatim}
\usepackage{graphicx}          % Include figure files
\usepackage{bm}                % bold math
\usepackage{hyperref}          % add hypertext capabilities
\usepackage[dvipsnames]{xcolor}
\begin{document}
%\linenumbers

\title{Flux Estimates and Detection Prospects for Lunar Antineutrinos}

\author{Hang Hu}%
\affiliation{Guangdong Ocean University, Zhanjiang, 524088, China}

\author{Yaping Cheng}
\email{applelovei@qq.com}
\affiliation{Beijing Institute of Spacecraft Environment Engineering, Beijing, 100094, China}

 \author{Wan-Lei Guo}%
 \email{guowl@ihep.ac.cn}
 \affiliation{Institute of High Energy Physics, Chinese Academy of Sciences,
 Beijing 100049, China}

\date{\today}% It is always \today, today,
             %  but any date may be explicitly specified

\begin{abstract}
The distribution of heat-producing elements (U, Th, K) within the Moon is critical for understanding its thermal evolution and formation history. Based on a refined lunar interior model, we calculate the lunar antineutrino fluxes from radioactive decays of these elements at two representative detector locations that bracket the expected signal intensity. The maximum flux is found to be slightly lower than the geoneutrino flux for the KamLAND site on Earth, while the minimum flux is approximately a factor of 9 lower than this maximum value. We also compute the expected event rates for lunar antineutrinos using three detection channels.

\end{abstract}

\maketitle

%%%%%%%%%%%%%%%%%%%%%%%%%%%%%%%%%%%%%%%%%%%%%%
%%%%%%%%%%%%%%%%%%%%%%%%%  Section I %%%%%%%%%
%%%%%%%%%%%%%%%%%%%%%%%%%%%%%%%%%%%%%%%%%%%%%%
\section{\label{sec1} Introduction}

The Moon exhibits a striking global dichotomy between its nearside and farside hemispheres, a feature that is fundamental to understanding planetary differentiation \cite{Laneuville2013}. The lunar surface is geochemically divided into three major provinces \cite{Jolliff2000}. The nearside is dominated by the Procellarum KREEP Terrane (PKT), a region characterized by extensive mare volcanism, high heat flow, and significantly enriched concentrations of heat-producing elements (HPEs) such as Uranium (U), Thorium (Th), and Potassium (K) \cite{Wieczorek2000, Shearer2006, Laneuville2018}. In contrast, the farside Feldspathic Highlands Terrane (FHT) is anorthositic, structurally thicker, and depleted in radioactive elements, representing the primordial lunar crust \cite{Ohtake2009, Wieczorek2013, ElkinsTanton2011}. A third distinct province, the South Pole-Aitken (SPA) basin, represents a massive impact structure that exposes deep crustal or upper mantle materials with unique compositional properties \cite{GarrickBethell2009, Moriarty2018}.

This asymmetry is fundamentally linked to the crystallization of the lunar magma ocean and the subsequent concentration of incompatible elements—Potassium, Rare Earth Elements, and Phosphorus (KREEP)—into the final residual melt \cite{Snyder1992}. The uneven distribution of this radiogenic KREEP layer is hypothesized to have driven the prolonged volcanic activity observed in the PKT \cite{Laneuville2013}. Understanding the precise 3D distribution of these elements is essential for constraining models of the Moon's formation and thermal history. While recent missions like Chang'e-5 have returned young basalts from the nearside \cite{Tian2021}, the recently completed Chang'e-6 mission has returned the first samples from the lunar farside, revealing distinct volcanic episodes at 2.8 Ga and 4.2 Ga that challenge previous thermal models \cite{Cui2024,Zhang2025}. Despite these advances, a global, deep-interior perspective remains necessary to resolve ambiguities regarding the bulk abundance of HPEs. Just as on Earth, the detection of lunar antineutrinos from radioactive decays of these elements offers a unique direct probe to constrain their distribution and advance our understanding of lunar evolution \cite{Fiorentini2007,Bellini2022,Dye2024}.

The radioactive decay of naturally occurring isotopes ($^{238}$U, $^{232}$Th, and $^{40}$K) is a primary heat source driving planetary thermal evolution \cite{Fiorentini2007,Bellini2022}. These decays emit electron antineutrinos $\bar{\nu}_e$, which traverse planetary bodies unimpeded, carrying information about the abundance and distribution of their parent isotopes. On Earth, the KamLAND experiment achieved the first detection of geoneutrinos in 2005, reporting a 90\% confidence interval of 4.5 to 54.2 events, which was consistent with geophysical predictions of 19 events \cite{Araki2005}. Subsequent advancements were marked by the Borexino, KamLAND, and SNO+ experiments, which observed geoneutrinos with high statistical significance. Borexino detected 53 events \cite{Agostini2020}, KamLAND 183 \cite{KamLAND:2022vbm}, and SNO+ 49~\cite{SNO+2025}, achieving relative precisions on the measured rate of 15\%, 18\%, and 23\%, respectively. In the near future, JUNO will measure the total geoneutrino flux with a precision of about 8\% over ten years \cite{JUNO:2025sfc}.

The Moon's internal heat sources remain poorly constrained. A roughly 9–13~mW/m$^2$ mantle heat flux best approximates the Apollo 15 and 17 observed heat fluxes~\cite{Siegler2014}, but the relative contributions from radiogenic decay, primordial accretion heat, and ongoing tidal interactions with Earth are actively debated among scientists. This uncertainty stems from limited direct sampling, variable crustal enrichment in heat-producing elements post-giant impact formation, and questions about the Moon's core and mantle dynamics \cite{Santangelo2025}. Recent research highlights the role of tidal heating, revealing an ultralow-viscosity zone at the core-mantle boundary where Earth's gravitational pull generates significant heat, suggesting the Moon's deep interior remains warm and partially molten rather than fully solidified as previously assumed \cite{Harada2014}. Extending this antineutrino detection technique to the Moon would allow for the measurement of the total lunar antineutrino flux, the refractory Th/U ratio, and the volatile K abundance. Such measurements would serve to experimentally verify these theoretical models. Furthermore, disentangling crustal and mantle contributions could resolve debates regarding the enrichment of the lunar mantle beneath the PKT, the overall bulk composition inherited from the giant impact hypothesis, and the thermal state of the core \cite{Jolliff2000, Weber2011}.

This study aims to estimate the lunar antineutrino fluxes on the Moon and calculate the expected event rates using diverse reaction channels. We utilize the updated lunar interior model to calculate fluxes at key locations. A major innovation of this work is the proposal to utilize in-situ lunar $^3$He resources as a target for detecting low-energy antineutrinos via radiochemical methods. We structure the paper as follows. Section~\ref{sec2} details the U, Th, and K distribution model. In Section~\ref{sec3}, we calculate the lunar antineutrino fluxes and angular distributions. Section~\ref{sec4} presents the expected event rates for three distinct reaction channels. Finally, a summary is given in Section~\ref{sec5}.

%%%%%%%%%%%%%%%%%%%%%%%%%%%%%%%%%%%%%%%%%%%%%%%%%%%%%%%
%%%%%%%%%%%%%%%%%%%%%%%%%  Section II  %%%%%%%%%%%%%%%%
%%%%%%%%%%%%%%%%%%%%%%%%%%%%%%%%%%%%%%%%%%%%%%%%%%%%%%%
\section{\label{sec2} Distributions of Th, U and K}

\begin{figure}
\centering
\includegraphics[width=0.450\textwidth]{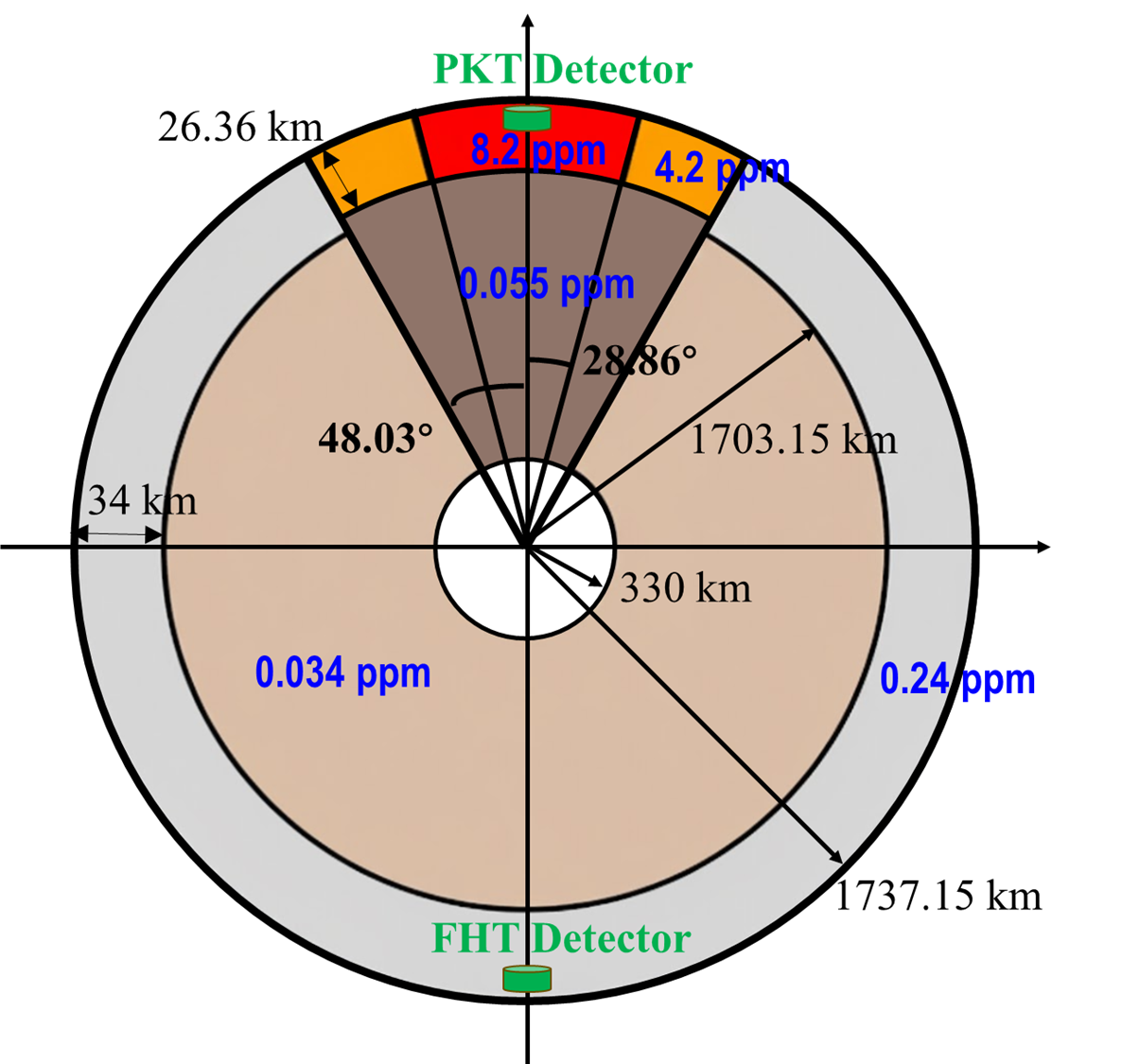}
\caption{Schematic cross-section of the lunar internal structure and the baseline Th distribution. The PKT crust consists of an inner region with 8.2 ppm Th and an outer region with 4.2 ppm Th. The PKT mantle (0.055 ppm Th) is slightly enriched compared to the highland mantle (0.034 ppm Th), while the FHT crust has 0.24 ppm Th. Two hypothetical neutrino detectors are placed in the PKT and FHT regions. }
\label{fig:Model5}
\end{figure}

To calculate the lunar antineutrino fluxes, a detailed understanding of the Moon's internal structure and the distribution of HPEs across its various reservoirs is required. The Moon exhibits a striking hemispheric dichotomy between the nearside and farside, which is reflected in its geochemical and crustal properties. Based on this asymmetry, the bulk silicate Moon (BSM) can be divided into five distinct geochemical reservoirs, each characterized by different HPE abundances \cite{Laneuville2018}: the PKT crust, the FHT crust, the SPA crust, the PKT mantle, and the FHT mantle. According to Ref.~\cite{Taylor2014}, the PKT crust covers 16.56\% of the lunar surface with an average thickness of 26.36 km, the FHT crust occupies 77.53\% with a thickness of 36.66 km, and the SPA crust accounts for 5.91\% with a thickness of 20.48 km. Following the approach of Ref.~\cite{Laneuville2018}, we simplify the Moon's internal structure by incorporating the SPA crust into the FHT crust, given its relatively small surface fraction, and assign the combined FHT crust a thickness of 34 km. Fig.~\ref{fig:Model5} shows the schematic cross-section of the simplified lunar internal structure, where the Moon is modeled as a sphere with a radius of 1737.15~km. Note that the PKT crust is divided into an inner region (angular extent 28.86$^\circ$ from the axis) and outer region (extending to 48.03$^\circ$), with the latter angular extent corresponding to the 16.56\% areal fraction of the PKT crust. This axisymmetric geometry captures the hemispheric asymmetry while simplifying the irregular PKT shape for calculations.

\begin{figure}
  \centering
  \includegraphics[width=0.50\textwidth]{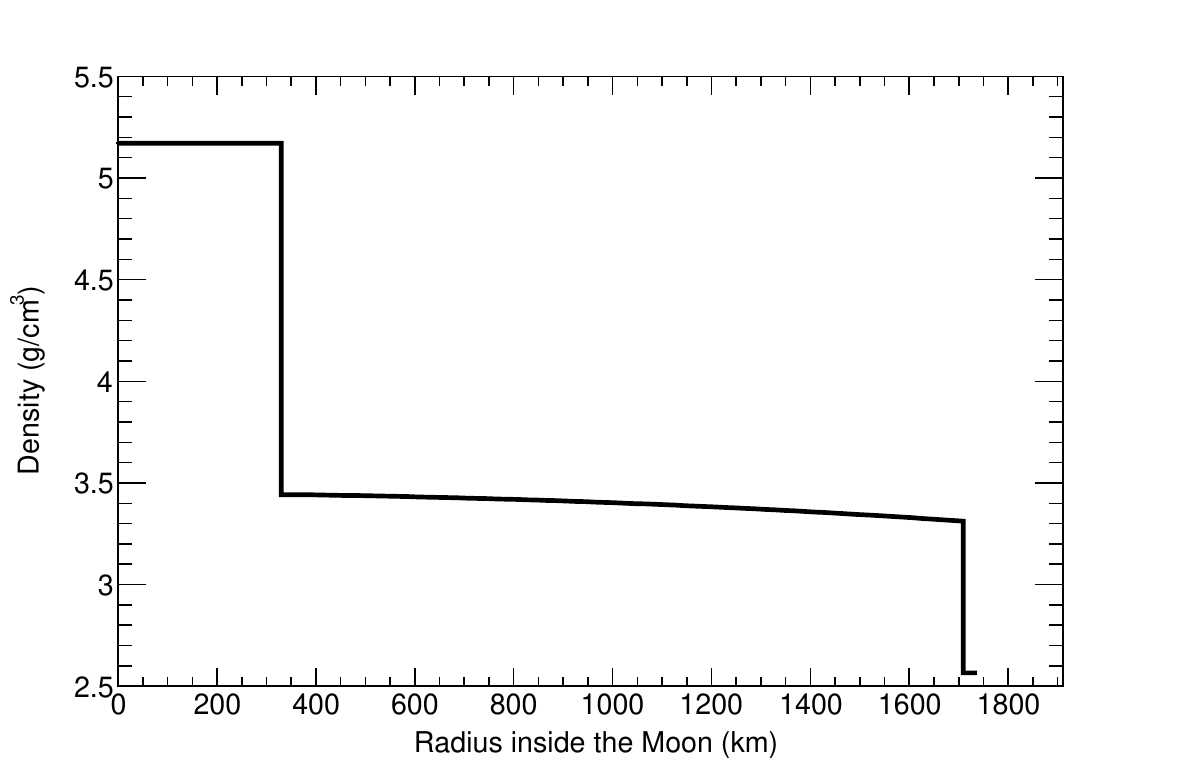}
  \caption{Lunar density versus radius.}
  \label{fig:density}
\end{figure}

For the lunar density profile, we use the results of Ref.~\cite{Garcia2011}. However, to ensure consistency with the crustal densities given in Ref.~\cite{Taylor2014}, the density of the uppermost 28 km layer is modified  to 2.566 g/cm$^3$. Furthermore, we adopt a core radius of 330 km, consistent with Ref.~\cite{Laneuville2018}. Combining the internal structure defined in Fig.~\ref{fig:Model5} with the modified lunar density profile shown in Fig.~\ref{fig:density}, we calculate the masses of the PKT inner crust, PKT outer crust, FHT crust, PKT mantle, and FHT mantle, as listed in Table~\ref{tab:masses}. The derived mass of the PKT crust ($4.18 \times 10^{20}\,\mathrm{kg}$) is consistent with the value of $4.22 \times 10^{20}\,\mathrm{kg}$ in Ref.~\cite{Taylor2014}. For the merged FHT crust (including the SPA crust), our value of $2.84 \times 10^{21}\,\mathrm{kg}$ agrees with the sum of the individual FHT and SPA crust masses from Ref.~\cite{Taylor2014}, $2.71 \times 10^{21}\,\mathrm{kg}$ and $0.118 \times 10^{21}\,\mathrm{kg}$, respectively.

\begin{table}[h]
\centering
\resizebox{\columnwidth}{!}{
\begin{tabular}{lccc}
\hline
Reservoir & Mass (kg)  & Mean Th (ppm) &  Th mass (kg)  \\
\hline
PKT Inner Crust & $1.57 \times 10^{20}$ & 8.2 & $1.3 \times 10^{15}$ \\
PKT Outer Crust & $2.61 \times 10^{20}$ & 4.2  & $1.1 \times 10^{15}$ \\
FHT Crust & $2.84 \times 10^{21}$ & 0.24 & $6.7 \times 10^{14}$ \\
PKT Mantle & $1.16 \times 10^{22}$ & 0.055  & $6.4 \times 10^{14}$ \\
FHT Mantle & $5.77 \times 10^{22}$ & 0.034 & $1.9 \times 10^{15}$ \\
\hline
Total & $7.26 \times 10^{22}$ & 0.0776 & $5.6 \times 10^{15}$ \\
\hline
\end{tabular}
}
\caption{Mass, mean Th abundance, and Th mass per reservoir for the baseline model.}
\label{tab:masses}
\end{table}

For the distribution of HPEs inside the Moon, Ref.~\cite{Laneuville2018} proposes five different models spanning a range of plausible scenarios. When tested with 3D thermochemical convection simulations, Model 5 best reproduces the key observables among the five candidate distributions, including the asymmetric and long-duration volcanism (lasting until $\sim$1-2 Ga on the nearside and $\sim$3 Ga on the farside), the paleomagnetic record of crustal magnetization, the Apollo heat flow measurements, and a sustained core dynamo for $\sim$1 Ga followed by episodic activity~\cite{Laneuville2018}. Here we adopt Model 5 as our baseline model for the calculation of lunar antineutrino fluxes. As shown in Fig.~\ref{fig:Model5}, Model 5 defines the Th distribution across five distinct geochemical reservoirs with the following concentrations~\cite{Laneuville2018}:
\begin{enumerate}
\item \textbf{PKT Inner Crust:} $8.2 \pm 4.0$ ppm Th. This enrichment corresponds to the composition of mafic impact-melt breccias and accounts for the observed lower magnetization in the PKT central region.
\item \textbf{PKT Outer Crust:} $4.2 \pm 2.5$ ppm Th. It is comparable to the lowest Th concentration traditionally used to define the PKT boundaries.
\item \textbf{FHT Crust:} $0.24 \pm 0.05$ ppm Th. This value is a composite that incorporates the upper 5~km aluminous layer ($1.0 \pm 0.3$  ppm), the underlying anorthosite ($0.05 \pm 0.03$ ppm), and the SPA crust ($1.5 \pm 0.4$~ppm).
\item \textbf{PKT Mantle:} $0.055 \pm 0.018$ ppm Th. It is derived from Th concentrations in mare basalts assuming melt fractions of $5\%$ and $10\%$, which give lower and upper bounds of $0.037$ ppm and $0.073$ ppm, and their mean is taken as the baseline value.
\item \textbf{FHT Mantle:} 0.034 ppm Th. This concentration is calculated from Th mass balance assuming a BSM composition analogous to the bulk silicate Earth (BSE), which differs from the 0.0366 ppm in Ref.~\cite{Laneuville2018} due to our use of the updated BSE Th abundance of 0.0776 ppm \cite{McDonough2026} instead of 0.0795 ppm \cite{McDonough1995}.
\end{enumerate}

It should be emphasized that the Th abundances adopted here are model-dependent and subject to considerable uncertainty, and will require sustained future investigation to be better constrained. In this work, we adopt Model 5 as a well-defined benchmark framework for calculating the resulting antineutrino fluxes, rather than treating it as a definitive compositional model of the Moon. An alternative perspective on the Moon’s composition has also been discussed in Ref. \cite{Siegler2022}. It is also worth noting that the BSE composition itself remains an active subject of debate \cite{McDonough2026}. Here we adopt the BSE Th abundance of 0.0776 ppm as a baseline to determine the FHT mantle Th abundance. The uncertainties in the Th abundances of the individual reservoirs, as well as the influence of alternative BSE Th values (0.0626–0.094 ppm) on our results, will be further discussed in Sec.~\ref{sec3}.

Using the baseline Th abundances and five reservoir masses, we calculate the Th mass for each reservoir, as listed in Table~\ref{tab:masses}. The U and K abundances can be derived from Th using fixed global ratios. We adopt a Th/U mass ratio of 3.77, following the empirical confirmation of Earth's chondritic Th/U from Ref.~\cite{McDonough2018}, as these refractory elements did not fractionate significantly during the Moon's formation via giant impact. The K abundance is derived using a bulk Moon K/Th ratio of 460~\cite{Laneuville2018}, reflecting the Moon's depletion in moderately volatile elements compared to Earth (where K/Th $\approx$ 3000 \cite{McDonough1995}) due to high-temperature processing and volatile loss in the protolunar disk.

This model was constructed by integrating multiple constraints: crustal thickness from GRAIL gravity data, surface Th abundances from Lunar Prospector gamma-ray spectrometer measurements, Apollo sample analyses, recent crustal magnetization inversions, partial melting estimates for mare basalt source regions, and mass balance considerations that adopt the commonly used working hypothesis of Earth-like bulk refractory element abundances inside the Moon (while recognizing that the HPE budget of the BSE itself remains debated). This model simplifies the PKT geometry into concentric cones and uses highland values to approximate the SPA basin properties, ensuring computational efficiency. Despite these approximations, it effectively captures the essential hemispherical asymmetry needed for a well-reasoned and representative flux estimation~\cite{Laneuville2018}. Additional simplifications include assuming uniform densities within the crust and mantle (ignoring minor lateral variations), a fixed core radius of 330~km without HPEs, and neglecting post-magma ocean processes like mantle overturn for initial condition setup.

%%%%%%%%%%%%%%%%%%%%%%%%%%%%%%%%%%%%%%%%%%%%%%%%%%%%%%%
%%%%%%%%%%%%%%%%%%%%%%%%%  Section III %%%%%%%%%%%%%%%%
%%%%%%%%%%%%%%%%%%%%%%%%%%%%%%%%%%%%%%%%%%%%%%%%%%%%%%%
\section{\label{sec3}  Lunar antineutrino fluxes }

The differential neutrino flux $\phi_i(E_{\bar{\nu}_e},\vec{d})$ at a detector position $\vec{d}$ can be calculated by integrating the contributions from all source volumes $V$ within the Moon. The flux formula is given by \cite{Fiorentini2007,Bellini2022}:
\begin{equation}
\phi_i(E_{\bar{\nu}_e},\vec{d}) = \int_V \frac{A_i(\vec{r}) ~ C_i }{\tau_i ~ m_i} f_i (E_{\bar{\nu}_e}) P_{ee} \frac{\rho(\vec{r})}{4\pi | \vec{r}-\vec{d} |^2} dV \;,
\label{eq:flux}
\end{equation}
where $A_i(\vec{r})$ is the abundance of the $i$th HPEs at position $\vec{r}$,  $C_i, \tau_i$ and $m_i$ are the isotopic concentration, lifetime and mass of nucleus $i$. The lunar density $\rho(\vec{r})$ can be obtained from Fig.~\ref{fig:density}.  $f_i (E_{\bar{\nu}_e})$ is the differential energy spectrum of the produced antineutrino per decay chain \cite{Enomoto}. We calculate the electron antineutrino survival probability $P_{ee}$ using the following formula:
\begin{align}
P_{ee} = \, & 1 - \sin^2 2\theta_{13} \left[ \cos^2\theta_{12} \cdot \Delta_{31} + \sin^2\theta_{12} \cdot \Delta_{32} \right]  \nonumber \\
& - \cos^4\theta_{13} \sin^2 2\theta_{12} \cdot \Delta_{21} \;,
\label{eq:pee}
\end{align}
with  $ \Delta_{ij}  = \sin^2 (1.27 \cdot |\Delta m^2_{ij}| \cdot | \vec{r}-\vec{d} | / E_{\bar{\nu}_e})$. 
For the neutrino oscillation parameters, assuming normal mass ordering, we take $\sin^2 2\theta_{13} = 0.0851$~\cite{DayaBay:2023prl}, $\sin^2 \theta_{12} = 0.3092$, $\Delta m_{21}^2 = 7.50 \times 10^{-5}\,\mathrm{eV^2}$~\cite{JUNO:2025first}, and $\Delta m_{31}^2 = 2.513 \times 10^{-3}\,\mathrm{eV^2}$ \cite{Esteban:2024eli}.  In the following calculation, $P_{ee}$ is evaluated numerically on a voxel-by-voxel basis: within 1 km of the detector, a step size of 1 m is used, while a coarser step of 100 m is adopted elsewhere, with $P_{ee}$ calculated at the center of each voxel. Eqs.~\eqref{eq:flux} and \eqref{eq:pee} account for the production rate of antineutrinos from beta decays in radiogenic isotopes  as they propagate through the lunar interior with negligible absorption.

We calculate the lunar antineutrino fluxes for two potential detector locations to bracket the signal intensity: the center of the PKT (maximum flux due to local crustal enrichment) and the center of the FHT (minimum flux), as shown in Fig.~\ref{fig:Model5}. The detector overburden of 50 m is chosen based on the typical depth of lunar lava tubes \cite{tubes}, which offer a naturally shielded environment for potential lunar bases and neutrino detectors. A distinct advantage of a lunar-based neutrino experiment at a depth of 50 m is the significant suppression of cosmic-ray backgrounds due to the lack of a lunar atmosphere.

\begin{figure}
  \centering    
  \includegraphics[width=0.5\textwidth]{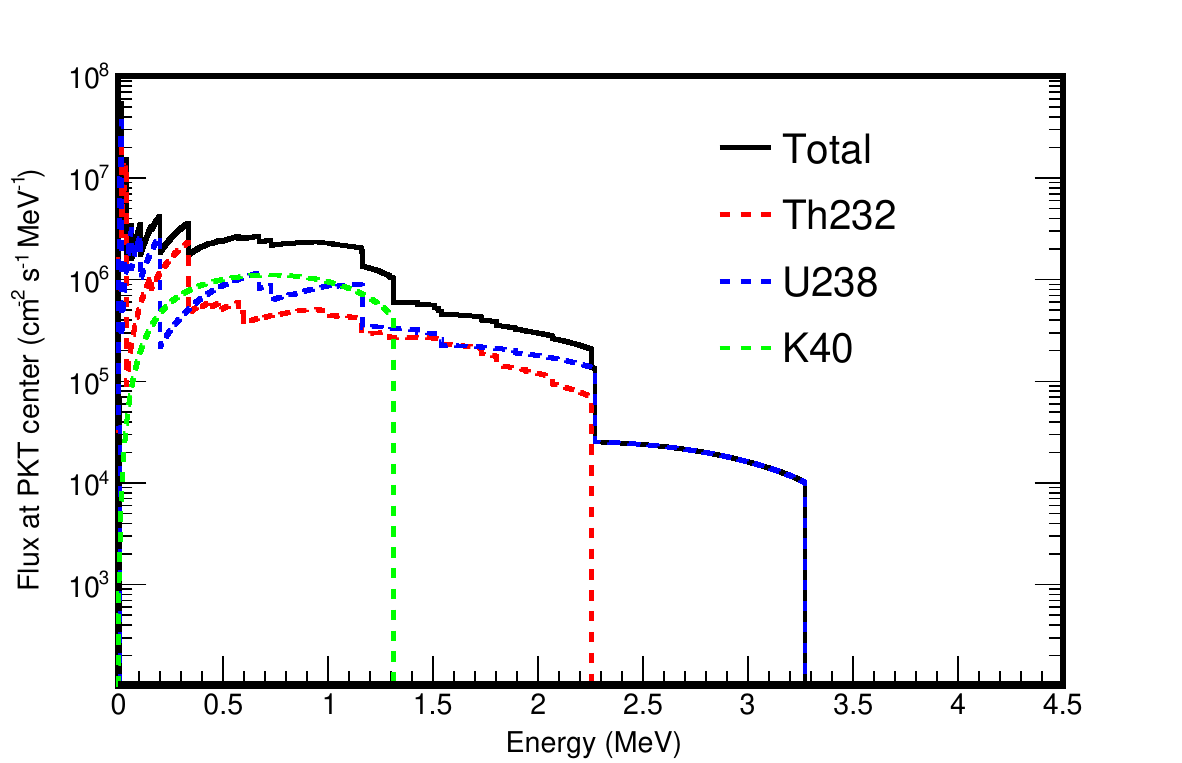}
  \caption{Lunar antineutrino energy spectra for the PKT detector, showing contributions from $^{232}$Th, $^{238}$U, and $^{40}$K. }
  \label{fig:flux}
\end{figure}

The antineutrino spectrum for the PKT detector is shown in Fig.~\ref{fig:flux}. The FHT spectrum can be obtained by scaling this curve down by a factor of $8.7$, which is the ratio of the integrated fluxes at the PKT and FHT sites. The integrated fluxes are determined to be $3.9 \times 10^6\,\mathrm{cm^{-2}\,s^{-1}}$ for the PKT detector and $4.4 \times 10^5\,\mathrm{cm^{-2}\,s^{-1}}$ for the FHT detector, respectively. The results obtained with the averaged survival probability $\langle P_{ee} \rangle \approx 0.5486$ differ from the above values by less than $4\%$. The PKT detector experiences the highest flux because it is situated directly above the region with the greatest concentration of HPEs, leading to enhanced local antineutrino production from the enriched inner and outer PKT crust. Conversely, the FHT detector registers the lowest flux as it overlies the feldspathic highlands with significantly lower radiogenic element abundances, resulting in a dominant contribution from the more distant and dilute mantle sources. For the PKT detector, the contributions from the  $^{238}$U and $^{232}$Th decay chains are calculated as $1.6 \times 10^6~ {\rm cm}^{-2} {\rm s}^{-1}$ and $1.3 \times 10^6~ {\rm cm}^{-2} {\rm s}^{-1}$. These values are slightly lower than the geoneutrino fluxes of $2.34\times 10^6~ {\rm cm}^{-2} {\rm s}^{-1}$ and $1.98\times 10^6~ {\rm cm}^{-2} {\rm s}^{-1}$ for the KamLAND site on Earth~\cite{Araki2005}.

\begin{figure}
  \centering
  \includegraphics[width=0.5\textwidth]{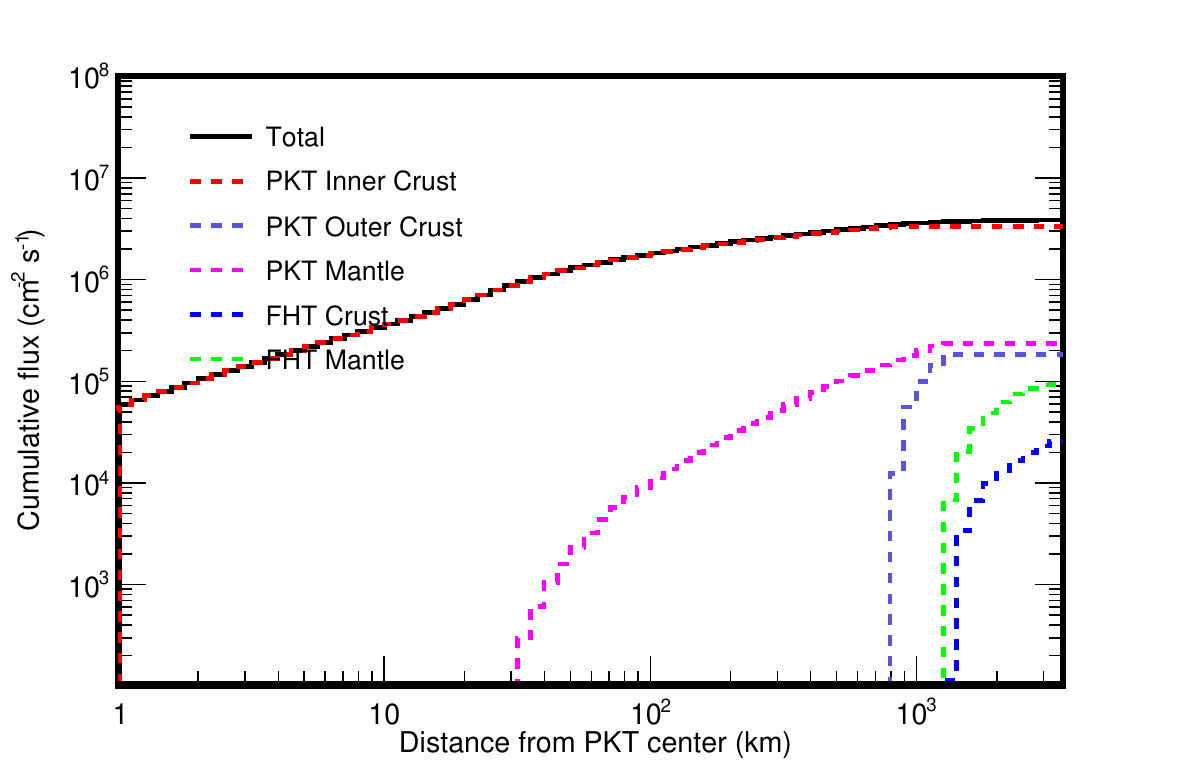}
  \includegraphics[width=0.5\textwidth]{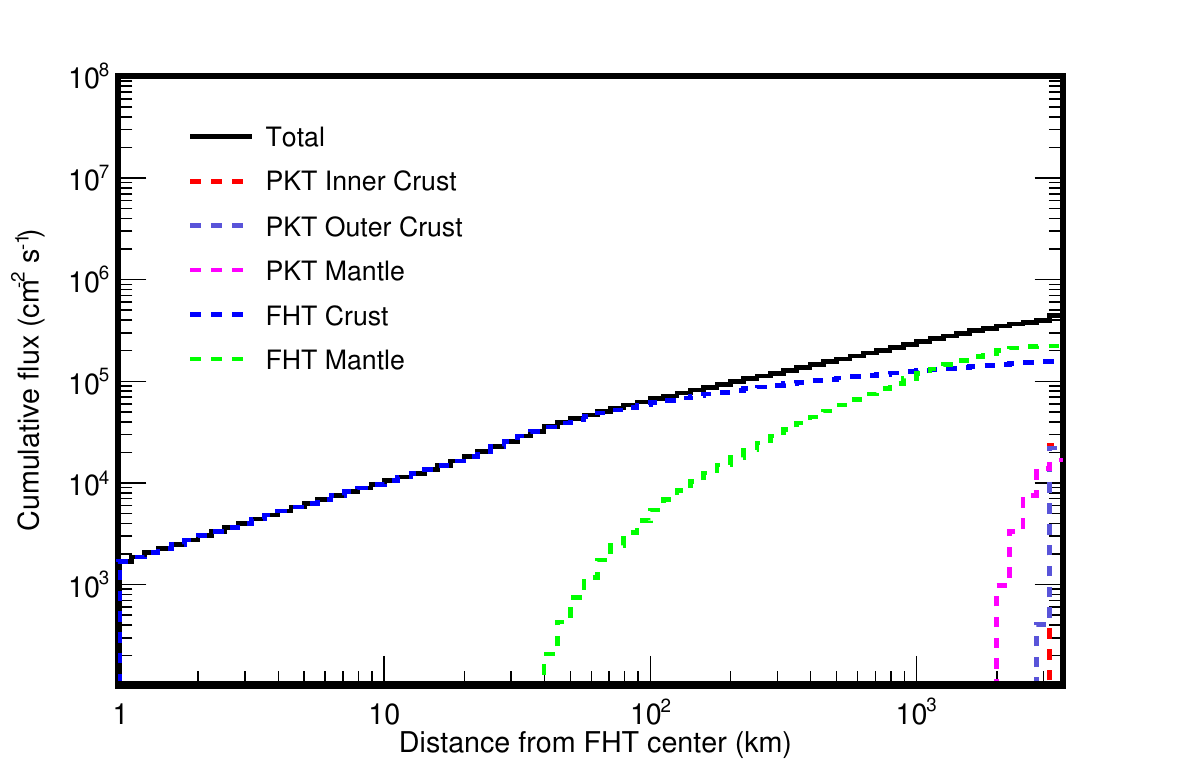}
  \caption{ Cumulative antineutrino fluxes as functions of the distance from the PKT  (top) and FHT (bottom) detectors, showing contributions from five distinct reservoirs.}
  \label{fig:distance}
\end{figure}

The cumulative neutrino fluxes as a function of the distance from the PKT and FHT detectors have been plotted in Fig.~\ref{fig:distance}. For the PKT case, the PKT inner crust contributes almost all the lunar antineutrinos within 1000~km. This rapid accumulation stems from the physics of source proximity: the PKT inner crust, with its high emissivity and lateral extent ($\sim$28.86° angular radius, corresponding to $\sim$862 km surface arc length at lunar radius 1710.79~km), encompasses most high HPE sources within intermediate distances. For the FHT detector, within 100 km, the FHT crust contributes most to the total signal, reflecting the immediate subsurface layer's role despite low HPEs. At a distance of 1000 km, the contributions from the FHT crust and FHT mantle become almost equal, as the cumulative integral incorporates extended crustal regions and the underlying mantle, whose larger homogeneous volume compensates for lower emissivity over mid-range distances. This parity highlights the transition from local crustal dominance to volumetric mantle influence in low HPE regions. These spatial patterns resonate with terrestrial geoneutrino observations \cite{Sramek2013, Fiorentini2005}, where cumulative fluxes at continental sites (analogous to the lunar PKT) receive approximately 50\% of their signal from within a 250 km radius due to the presence of thick, radioelement-enriched crust. In contrast, oceanic or thin-crust sites (analogous to the FHT) is less localized and features a non-negligible far-field mantle component. The Moon’s compact geometry amplifies the angular resolution of crustal heterogeneities like the PKT, yet simultaneously demands directional detection to partition the superimposed signals from layered interior reservoirs.

\begin{figure}
  \centering
  \includegraphics[width=0.5\textwidth]{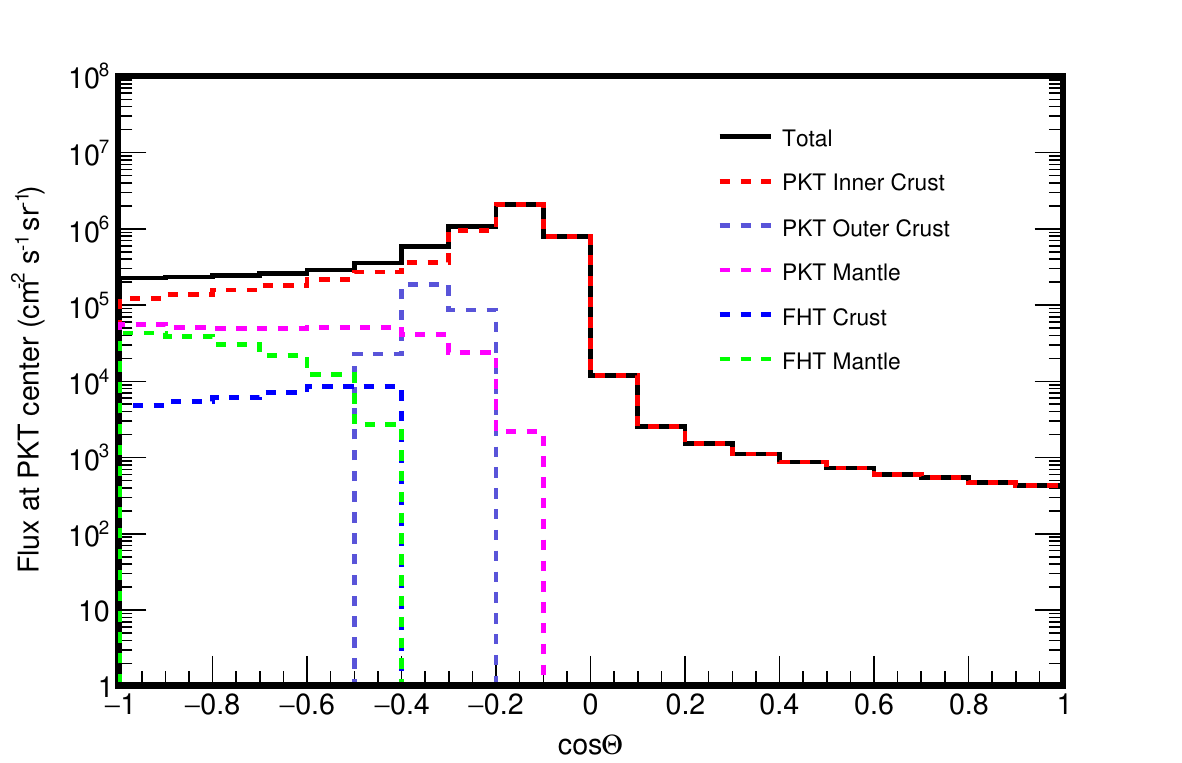}
  \includegraphics[width=0.5\textwidth]{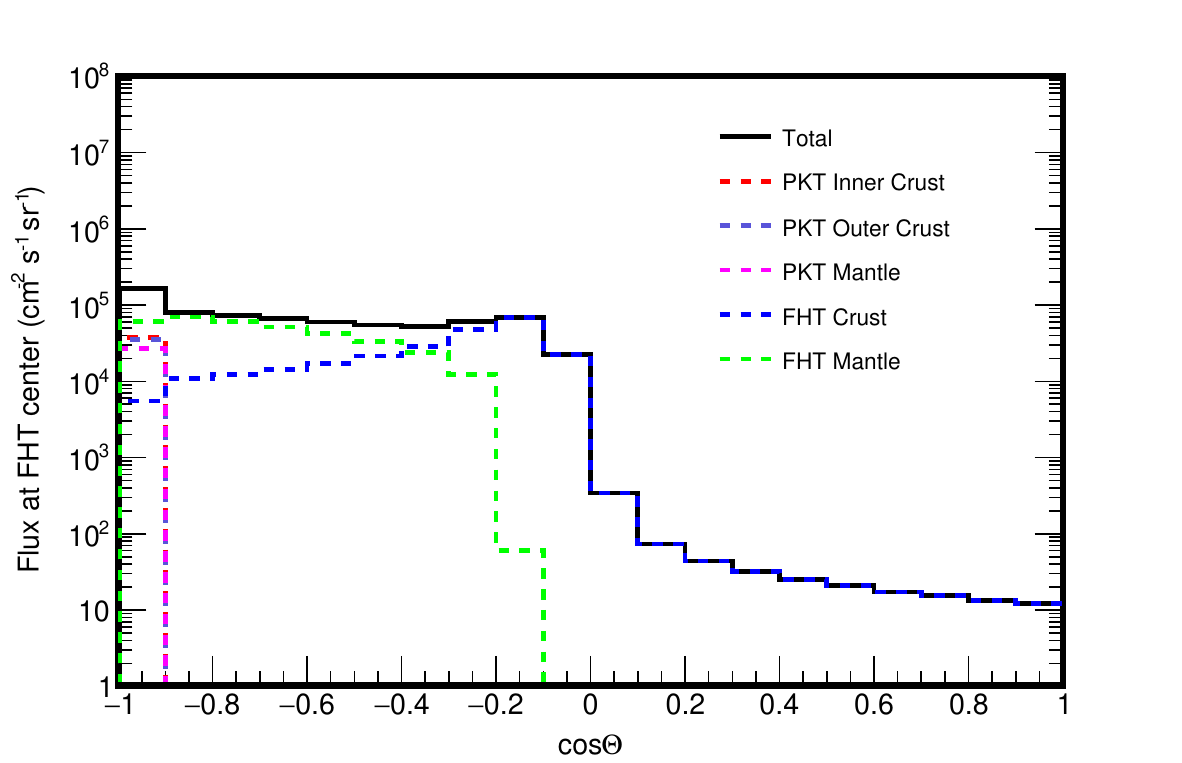}
  \caption{Angular distributions of lunar antineutrinos at the PKT (top) and FHT (bottom) detectors  from five reservoirs.}
  \label{fig:direction}
\end{figure}

The capability to measure the direction of incident neutrinos would provide significant benefits. Directional sensitivity allows separation of the FHT crust and mantle signals, as shown in Fig.~\ref{fig:direction}. The FHT crust antineutrinos peak at near-horizontal incident angles ($\cos \theta = 0$), while the FHT mantle antineutrinos would be dominant over a range of angles below the horizontal ($\cos \theta < -0.4$). Directional detection would enable a tomography of the lunar interior. For the PKT detector, the PKT outer crust can give a significant contribution in the range of $ -0.4 < \cos \theta < -0.3$. Directional detection of antineutrinos from this angular range can help constrain the boundaries of the enriched inner PKT region, thus verifying heat source distribution models \cite{Laneuville2018}. Furthermore, this directionality aids in subtracting residual backgrounds (e.g., solar and potential reactor neutrinos) and could probe deeper structures, such as core contributions or asymmetries in HPEs distribution, as demonstrated in terrestrial studies with direction-sensitive detectors \cite{Leyton2017, Tanaka2014}.

\begin{figure}
  \centering
  \includegraphics[width=0.5\textwidth]{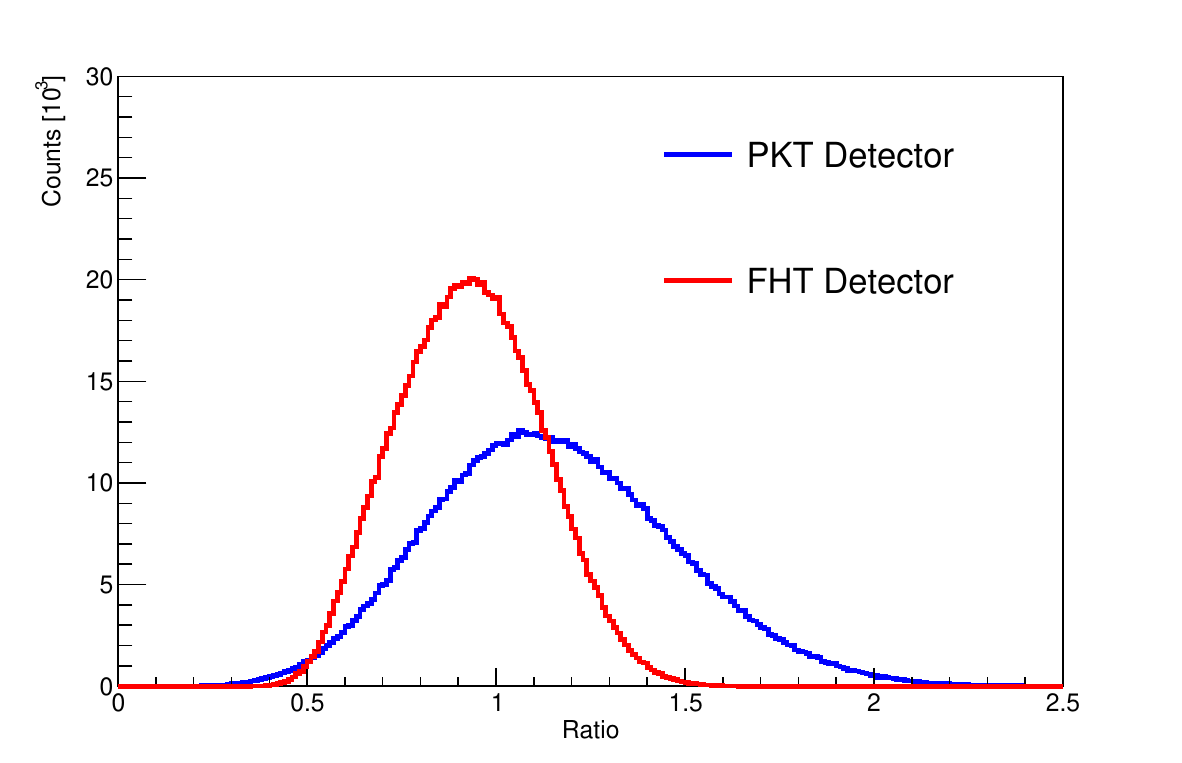}
  \caption{Distributions of the predicted lunar antineutrino fluxes at the PKT and FHT sites, normalized to the baseline values, showing the impact of Th abundance uncertainties.}
  \label{fig:sample}
\end{figure}

All flux results presented above are based on our baseline model, which does not incorporate uncertainties in the Th abundances of the individual reservoirs. To assess the impact of these uncertainties on the antineutrino fluxes, we sample the Th abundance of each reservoir according to its uncertainty, with the FHT mantle abundance determined via mass balance adopting a BSE Th abundance of 0.0776 ppm. Fig.~\ref{fig:sample} shows the predicted flux distributions normalized to the baseline values of $3.9 \times 10^6\,\mathrm{cm^{-2}\,s^{-1}}$ (PKT) and $4.4 \times 10^5\,\mathrm{cm^{-2}\,s^{-1}}$ (FHT), with each distribution based on $10^6$ random samplings. The peaks of both distributions slightly deviate from unity. This is because we impose the constraints that the sampled Th abundances must satisfy PKT inner crust $>$ PKT outer crust $>$ FHT crust $>$ 0 and PKT mantle $>$ FHT mantle $>$ 0. It is clear that the majority of the predicted fluxes for both detectors fall within the range of $0.5$ to $2.0$ times the baseline values. This can be understood from Fig.~\ref{fig:distance}, where the PKT flux is predominantly determined by the Th abundance of the PKT inner crust ($8.2 \pm 4.0$ ppm), while for the FHT detector, nearly half of the flux originates from the FHT crust, whose uncertainty is relatively small ($0.24 \pm 0.05$~ppm). 

The FHT flux is primarily controlled by the Th abundance of the FHT mantle, which is derived from mass balance adopting a BSE Th abundance of $0.0776$ ppm. However, the BSE Th abundance remains debated in the geochemical community \cite{McDonough2026}. To quantify this dependence, we vary the BSE Th abundance to $0.0626$ ppm \cite{Lyubetskaya2007} and $0.094$~ppm \cite{Jagoutz1979} while keeping the other four reservoir abundances fixed at their baseline values. Mass balance then yields FHT mantle Th abundances of $0.015$ ppm and $0.054$ ppm, respectively. The resulting FHT fluxes are $3.2 \times 10^5\,\mathrm{cm^{-2}\,s^{-1}}$ and $5.8 \times 10^5\,\mathrm{cm^{-2}\,s^{-1}}$, respectively, corresponding to deviations of about $\pm 30\%$ from the baseline value of $4.4 \times 10^5\,\mathrm{cm^{-2}\,s^{-1}}$. In contrast, the PKT flux shows little variation, changing from $3.8\times 10^6$ to $3.9 \times 10^6\,\mathrm{cm^{-2}\,s^{-1}}$ for the two models.

%%%%%%%%%%%%%%%%%%%%%%%%%%%%%%%%%%%%%%%%%%%%%%%%%%%%%%
%%%%%%%%%%%%%%%%%%%%%%%%% Section IV %%%%%%%%%%%%%%%%%%
%%%%%%%%%%%%%%%%%%%%%%%%%%%%%%%%%%%%%%%%%%%%%%%%%%%%%%
\section{\label{sec4} Event rate estimates for lunar antineutrino detection}

In this section, we shall estimate the expected event rates of lunar antineutrinos through three distinct interaction channels: 
\begin{enumerate}
\item $\bar{\nu}_e + p \to e^+ + n$ (Inverse Beta Decay, IBD);
\item $\bar{\nu}_e + ^3$He $\to e^+ + ^3$H (IBD reaction on $^3$He);
\item $\bar{\nu} + e^- \to \bar{\nu} + e^-$ (Elastic Scattering, ES).
\end{enumerate}

\subsection{\label{sec4.1} IBD reaction on free protons }

\begin{figure}
  \centering
  \includegraphics[width=0.5\textwidth]{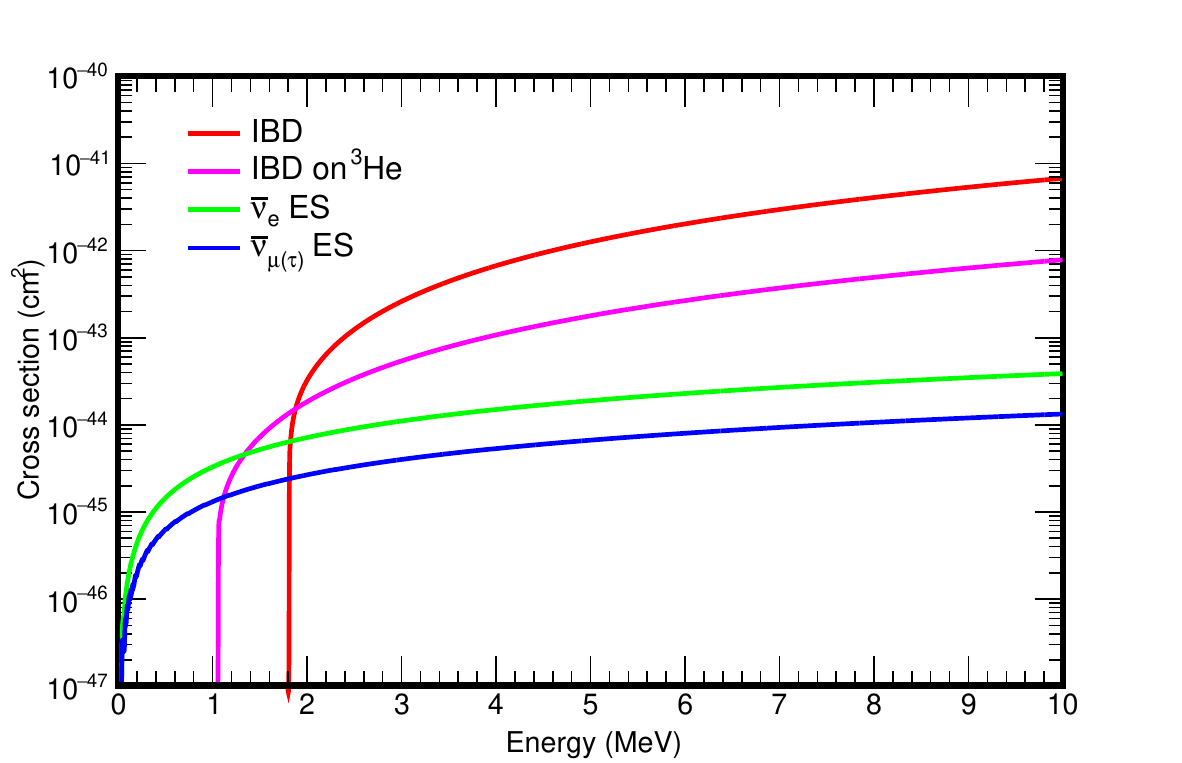}
  \caption{Antineutrino cross sections for three interaction channels: IBD reaction with 1.806 MeV threshold, IBD reaction on $^3$He with 1.041 MeV threshold and elastic scattering of $\bar{\nu}_e$ and $\bar{\nu}_{\mu (\tau)}$ on electrons with no energy threshold.}
  \label{fig:CS}
\end{figure}

\begin{figure}
  \centering
  \includegraphics[width=0.5\textwidth]{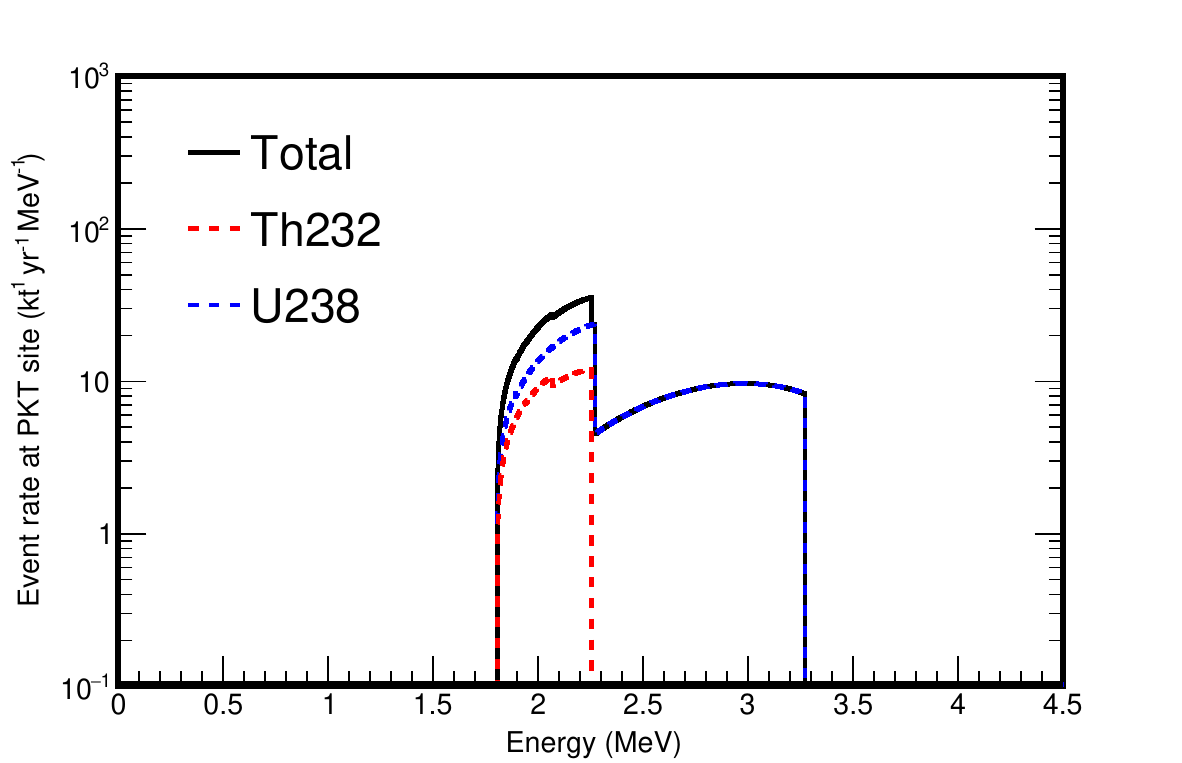}
  \caption{IBD event rates from $^{238}$U and $^{232}$Th at the PKT site.}
  \label{fig:IBD}
\end{figure}

The IBD reaction, $\bar{\nu}_e + p \to e^+ + n$, serves as the primary channel for geoneutrino detection on Earth due to its interaction cross section \cite{Strumia:2003zx}, as shown in Fig.~\ref{fig:CS}, and the powerful background rejection offered by its distinct prompt-delayed coincidence signature in the liquid scintillator detectors \cite{Araki2005,Agostini2020,KamLAND:2022vbm,SNO+2025,JUNO:2025sfc,Jinping:2016iiq}. However, a fundamental limitation of this channel is its energy threshold of 1.806 MeV. This threshold renders the reaction sensitive exclusively to antineutrinos originating from the decay chains of $^{238}$U and $^{232}$Th, while leaving the detector blind to the flux from $^{40}$K decays, which possess a maximum energy of approximately 1.31~MeV. Assuming a liquid scintillator detector composed of 88\% $^{12}$C and 12\% $^{1}$H by weight \cite{JUNO:2015zny}, we calculate the expected IBD event rates for the PKT site, as shown in Fig.~\ref{fig:IBD}. The total rate is found to be 19 kt$^{-1}$ yr$^{-1}$, where $^{238}$U contributes 15~kt$^{-1}$~yr$^{-1}$.  The rates for the FHT detector are lower by a factor of 8.7. For comparison, the Earth's geoneutrino luminosity reported in Ref.~\cite{McDonough2020} would yield only about $0.001 \,\mathrm{kt^{-1}\,yr^{-1}}$, which is entirely negligible relative to the lunar signal.

\subsection{\label{sec4.2} IBD reaction on $^3$He }

Here we utilize $^3$He as the target nucleus to detect the lunar antineutrinos through the charged current reaction: 
\begin{equation} 
\bar{\nu}_e + ^3{\rm He} \to e^+ + ^3{\rm H}\;. 
\end{equation} 
This channel offers several significant advantages. Firstly, this reaction possesses a comparatively large cross section \cite{Golak:2018qya}, as shown in Fig.~\ref{fig:CS}. Note that the cross section used here is derived from high precision nuclear theory calculation, specifically utilizing the weak charged single-nucleon current operator and the AV18 potential for the $^3$He and $^3$H wave functions \cite{Golak:2018qya}. Secondly, the reaction product is $^3$H, which has a half-life of 12.32 years, allowing for accumulation over time. Most importantly, the threshold for this reaction is 1.041 MeV. Unlike the IBD threshold, this lower threshold allows for the detection of the high energy tail of $^{40}$K antineutrinos (endpoint energy 1.31 MeV). Indeed, this reaction was discussed long ago for the detection of terrestrial geoneutrinos \cite{Krauss:1983zn,Kobayashi1991}. However, it has remained experimentally impractical due to the unlikelihood of obtaining sufficient quantities of $^3$He on Earth. Building on this limitation, the approach proposed here leverages the unique lunar environment, where the total inventory of $^3$He amounts to about $6.5 \times 10^5$ ton in the regolith due to solar wind implantation \cite{Fa2007}, constituting a valuable resource for future fusion power and in-situ resource utilization. 

\begin{figure}
  \centering
  \includegraphics[width=0.5\textwidth]{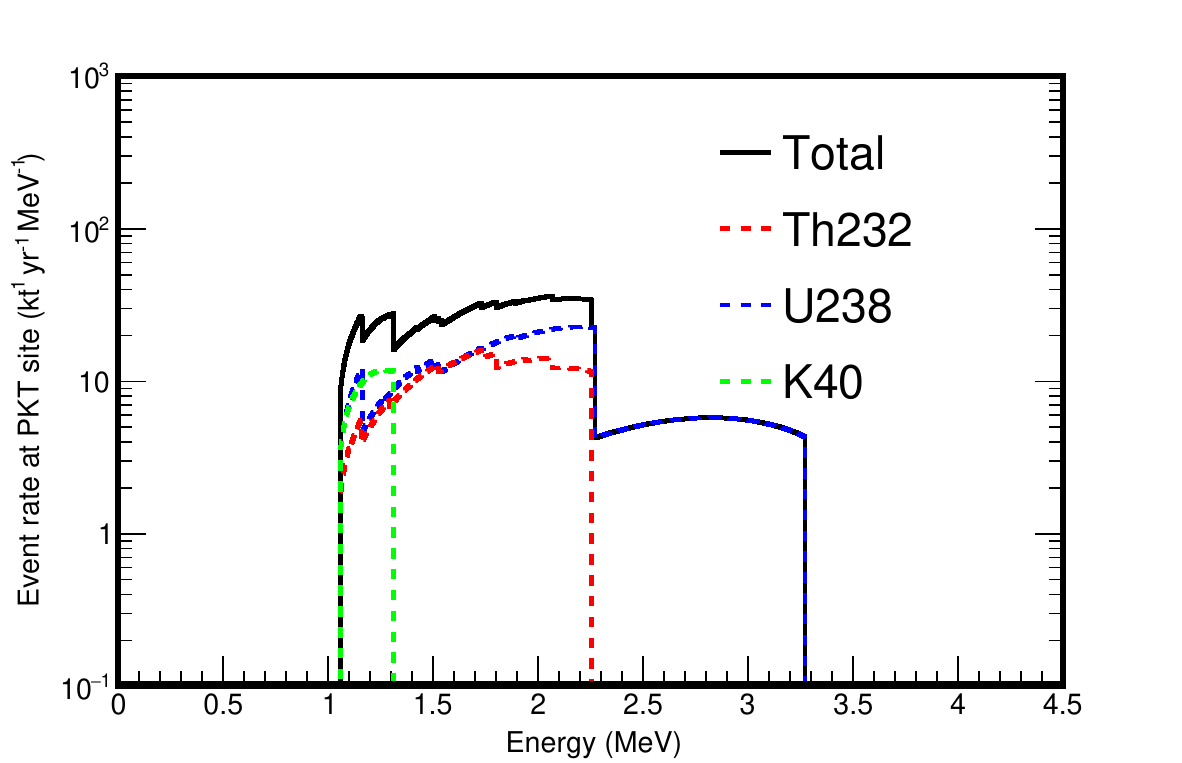}
  \caption{Event rates of the IBD reaction on $^3$He from $^{232}$Th, $^{238}$U and $^{40}$K at the PKT site.}
  \label{fig:He3}
\end{figure}

We propose a radiochemical method to measure lunar antineutrinos, mirroring the pioneering Homestake chlorine detector that used the reaction ${\nu}_e + ^{37}$Cl $\to e^- + ^{37}$Ar~\cite{Cleveland:1998nv}. In this scenario, $^3$H generated by the reaction $\bar{\nu}_e + ^3$He $\to e^+ + ^3$H undergoes simultaneous decay. Therefore, its net inventory over time is given by
\begin{equation} 
N(t) = N_{\rm eq} \times (1 - e^{- \lambda \cdot t}) \;, 
\end{equation} 
where the decay constant $\lambda = 0.0563 \, \text{yr}^{-1}$ and  $t$ is the exposure time. For a given event rate $R$, the equilibrium number of $^3$H is $N_{\rm eq} = R/\lambda$. This substantial ratio indicates a strong accumulation effect relative to the instantaneous production rate. To estimate the accumulation number of $^3$H atoms, we calculate the expected event rate for the PKT site, as shown in Fig.~\ref{fig:He3}. It indicates that $^{40}$K provides a non-negligible contribution to the event rate in the low energy region. The total production rate $R$ of $^3$H is found to be 39 kt$^{-1}$ yr$^{-1}$. For illustration, we consider a 1 kt $^3$He target with a 10-year exposure, yielding a net $^3$H production of approximately $3.0\times 10^2$ atoms, where the contributions from $^{238}$U, $^{232}$Th, and $^{40}$K are  $1.8\times 10^2$,  $1.0\times 10^2$, and  $1.9\times 10^1$ atoms, respectively. The practical implementation of this radiochemical method faces many technical challenges, such as two dangerous backgrounds: $n + {}^3\mathrm{He} \rightarrow p + {}^3\mathrm{H}$ and $\mu^- + {}^3\mathrm{He} \rightarrow \nu_\mu + {}^3\mathrm{H}$, with both neutrons \cite{Naito} and stopped muons largely sourced from cosmic-ray interactions. Addressing these challenges requires dedicated research that extends beyond the focus of the present work.

\subsection{\label{sec4.3} Elastic scattering on electrons }

\begin{figure}
  \centering
  \includegraphics[width=0.5\textwidth]{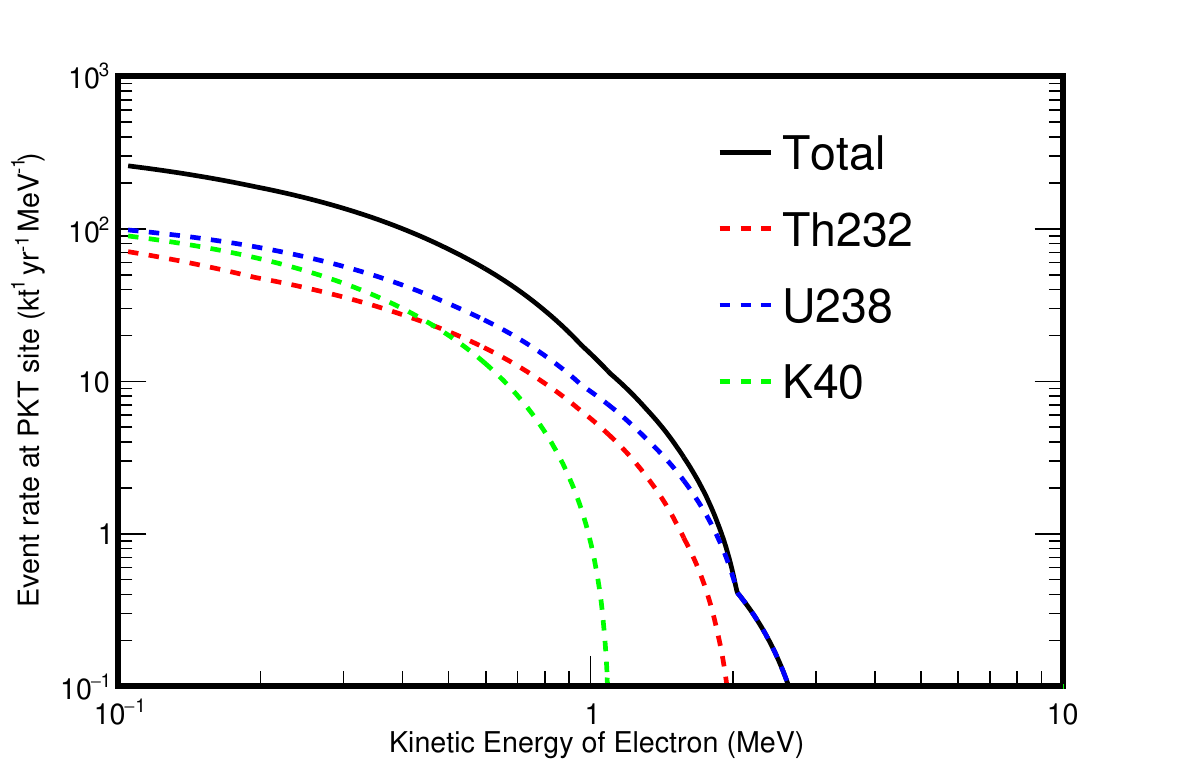}
  \caption{Kinetic energy spectra of recoiling electrons produced by the ES reaction of lunar antineutrinos for the PKT detector.}
  \label{fig:ES}
\end{figure}

The third detection channel is neutrino-electron elastic scattering, $\bar{\nu} + e^- \to \bar{\nu} + e^-$. This process has no intrinsic energy threshold, making it sensitive to the full lunar antineutrino spectra—a particularly important feature for detecting lunar antineutrinos from $^{40}$K. It should be noted that $\bar{\nu}_e$, $\bar{\nu}_\mu$ and $\bar{\nu}_\tau$  all participate in elastic scattering with electrons, with their respective cross sections shown in Fig.~\ref{fig:CS} \cite{Giunti}. In Fig.~\ref{fig:ES}, we plot the kinetic energy spectra of recoiling electrons produced by the elastic scattering reaction of lunar antineutrinos for the PKT detector. The expected overall elastic scattering event rate is found to be $1.2 \times10^2$~kt$^{-1}$~yr$^{-1}$, with the contributions coming from $^{238}$U, $^{232}$Th, and $^{40}$K given by 49, 33, and 34 kt$^{-1}$ yr$^{-1}$, respectively. This rate already exceeds the IBD event rate, despite the fact that the ES cross sections are much smaller. The higher rate is attributed to a far larger number of target electrons, low energy lunar antineutrinos below the IBD threshold ($<$1.806~MeV), and the additional contribution from $\bar{\nu}_\mu$ and $\bar{\nu}_\tau$. However, not all ES events are suitable for $^{40}$K analysis because solar neutrinos constitute a severe background. In the relevant energy range, the solar neutrino flux exceeds the lunar antineutrino flux by orders of magnitude, especially for $pp$ chain neutrinos. Directional reconstruction and an energy threshold can help to reduce this background. Following Ref.~\cite{Wang:2017etb}, we adopt a threshold of $0.7\,\mathrm{MeV}$ for the recoil electron kinetic energy, reducing the overall signal rate to $13\,\mathrm{kt^{-1}\,yr^{-1}}$. Even in this case, the signal rate remains far below the solar neutrino background~\cite{Wang:2017etb}. Directional reconstruction could further suppress this background, but efforts to achieve such directional detection have so far not succeeded.

%%%%%%%%%%%%%%%%%%%%%%%%%%%%%%%%%%%%%%%%%%%%%%%%%%%%%%%
%%%%%%%%%%%%%%%%%%%%%%%%%  Section V %%%%%%%%%%%%%%%%%%
%%%%%%%%%%%%%%%%%%%%%%%%%%%%%%%%%%%%%%%%%%%%%%%%%%%%%%%

\section{\label{sec5} Summary}

In summary, we have derived comprehensive flux estimates for lunar antineutrinos using a refined interior model for the distribution of heat-producing elements in the lunar crust and mantle, focusing on the PKT and FHT as high- and low-flux benchmarks, respectively. The integrated fluxes are determined to be $3.9 \times 10^6\,\mathrm{cm^{-2}\,s^{-1}}$ for the PKT detector and $4.4 \times 10^5\,\mathrm{cm^{-2}\,s^{-1}}$ for the FHT detector. The high flux at the PKT originates almost exclusively from its inner crust, which hosts the highest abundances of heat-producing elements, whereas the low flux at the FHT arises from comparable contributions from both its crust and mantle. Accounting for the uncertainties in the Th abundances, most predicted fluxes for both detectors lie within a factor of two of the baseline values. We have also computed the angular distributions of lunar antineutrinos and shown that directional detection could, in principle, resolve the contributions from the FHT crust and mantle. Finally, the expected event rates for IBD, IBD on $^3$He, and ES at the PKT detector are $19$, $39$, and $13\,\mathrm{kt^{-1}\,yr^{-1}}$ (the last with a $0.7\,\mathrm{MeV}$ threshold), respectively. The corresponding rates for the FHT detector are obtained by dividing these values by a factor of $8.7$. Ultimately, a lunar antineutrino observatory would provide a transformative, non-invasive probe of the Moon's deep interior, helping to resolve longstanding questions regarding its formation, differentiation, and thermal evolution.

\acknowledgments
The authors sincerely thank Prof. W. F. McDonough for his critical review, which greatly improved the treatment of HPE uncertainties and compositional models in this work. W. L. Guo is grateful to Prof. J. Golak for sharing the precise data for the cross sections of $\bar{\nu}_e$ on $^3$He in the 1.06–10 MeV range. This work is supported in part by the National Natural Science Foundation of China (NSFC) under Grants No. 12205018, 12375098 and U2541290, the National Key Laboratory of Reliability and Environmental Engineering Technology under Grant No. WDZC61420040202, and the Strategic Priority Research Program of the Chinese Academy of Sciences under Grant No. XDA10010100.

\end{document}